\documentclass[format=acmsmall, review=false, screen=true,  nonacm=true]{acmart}
\usepackage{booktabs} 

\usepackage[ruled]{algorithm2e} 

\SetAlFnt{\small}
\SetAlCapFnt{\small}
\SetAlCapNameFnt{\small}
\SetAlCapHSkip{0pt}
\IncMargin{-\parindent}

\usepackage{soul}
\usepackage{multirow}
\usepackage{multicol}

\begin{document}
\title[Vesyla]{Vesyla-II: An Algorithm Library Development Tool for Synchoros VLSI Design Style}
\author{Yu Yang}
\orcid{1234-5678-9012-3456}
\affiliation{%
  \institution{KTH Royal Institute of Technology}
  \streetaddress{Kistag\aa ngen 16}
  \city{Stockholm}
  \postcode{16440}
  \country{Sweden}}
\email{yuyang2@kth.se}

\author{Ahmed Hemani}
\orcid{1234-5678-9012-3456}
\affiliation{%
  \institution{KTH Royal Institute of Technology}
  \streetaddress{Kistag\aa ngen 16}
  \city{Stockholm}
  \postcode{16440}
  \country{Sweden}}
\email{hemani@kth.se}

\renewcommand{\shortauthors}{Y. Yang et al.}
\begin{abstract}
High-level synthesis (HLS) has been researched for decades and is still limited to fast FPGA prototyping and algorithmic RTL generation. A feasible end-to-end system-level synthesis solution has never been rigorously proven. Modularity and composability are the keys to enabling such a system-level synthesis framework that bridges the huge gap between system-level specification and physical level design. It implies that 1) modules in each abstraction level should be physically composable without any irregular glue logic involved and 2) the cost of each module in each abstraction level is accurately predictable. The ultimate reasons that limit how far the conventional HLS can go are precisely that it cannot generate modular designs that are physically composable and cannot accurately predict the cost of its design. In this paper, we propose Vesyla-II, not as yet another HLS tool, but as a synthesis tool that positions itself in a promising end-to-end synthesis framework and preserving its ability to generate physically composable modular design and to accurately predict its cost metrics. We present in the paper how Vesyla-II is constructed focusing on the novel platform it targets and the internal data structures that highlights the uniqueness of Vesyla-II. We also show how Vesyla-II will be positioned in the end-to-end synchoros synthesis framework called SiLago.

\end{abstract}

\maketitle
\section{Introduction}
\label{sec:intro}

High-level synthesis (HLS) has been researched for decades. Initially, many academic and industrial researchers wished to use HLS to massively accelerate the chip design process by enabling an automatic end-to-end (from high-level model to silicon layout) synthesis flow. However, the high-level synthesis (HLS) focuses on a single algorithm and synthesizes the algorithm in terms of basic micro-architectural building blocks such as adders, and multipliers. The current conventional HLS is not very satisfactory because its building foundation prevents it from further progress. An automatic end-to-end synthesis flow will require that on each abstraction level, the solutions are modular and composable so that higher abstraction design can be constructed by combining lower-level pieces. A good combining process is very difficult to achieve because it forbids the usage of irregular and unpredictable glue logic and requires that the cost of higher-level design can be accurately inferred by the cost of its lower level pieces. For example, the combining process used by conventional HLS tools is bad because it uses soft-IPs as building blocks whose cost is inaccurate, and it requires synthesized arbitrary glue logic to combine these soft-IPs to produce an RTL solution. Therefore, the RTL solutions emitted by HLS tools are not physically composable and the cost estimation of such solutions is not accurate at all because it needs further synthesis steps to obtain accurate estimation. These are the fundamental reasons that prevent conventional HLS tools from enabling application-level synthesis to automatically generate a more complex design. In other words, conventional HLS is stuck with synthesizing individual algorithms. Further improvement for automatically handling more complex synthesis of applications (such as wireless communication) is not possible.

A good HLS tool, on the other hand, should position itself in the centre of an end-to-end synthesis flow. It should target a composable physically hardened hardware platform and produces a compostable algorithmic solution whose cost metrics can be accurately predicted. The goal of such an HLS tool is to swiftly generate a Pareto set of algorithm implementations to form a library that enables more complex application-level synthesis (ALS).

To maintain the modularity, composability and reusability, the platform such HLS tool targets should be reconfigurable and efficient. The best strategy is to design such a platform to be application domain specific which could retain both properties. Thus, in an end-to-end synthesis flow, there would be many such hardware platforms and their corresponding HLS synthesis tools, each of which targets a specific application domain, such as dense linear algebra, wireless communication, graph theory, etc.

\begin{figure}[h]
    \centering
    \includegraphics[width=.9\textwidth, trim=.01cm .01cm .01cm .01cm, clip]{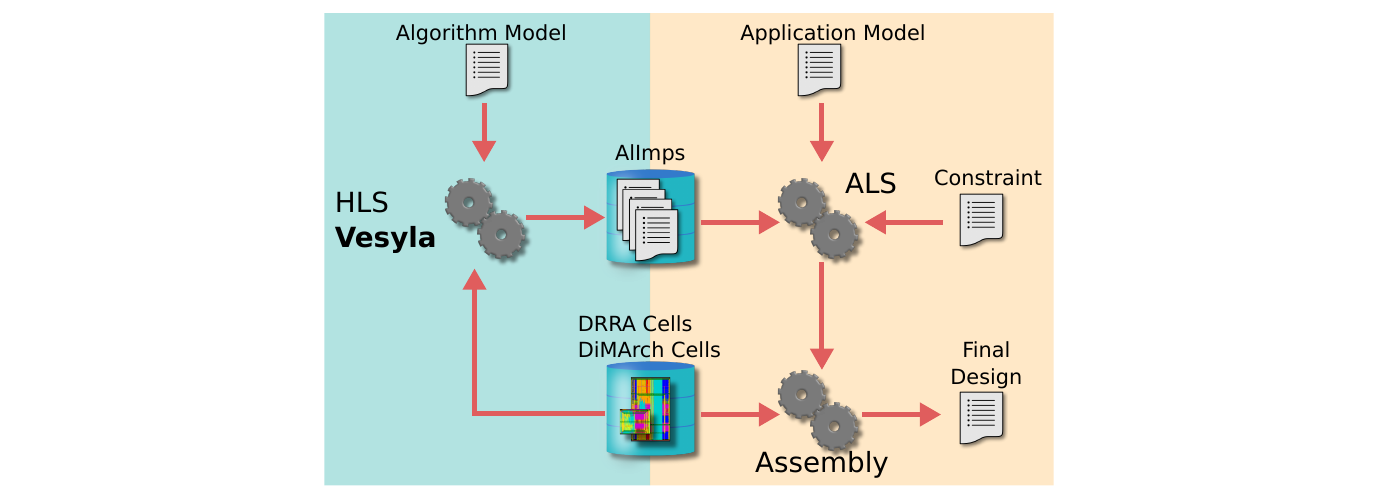}
    \caption{Relation between HLS and ALS in SiLago synthesis flow for streaming applications. The HLS tool Vesyla-II creates a library for algorithm implementations (AlImps) and the ALS uses such a library to synthesize applications. Based on \cite{Stathis2022SynchorosStyle}}
    \label{fig:als}
\end{figure}

Vesyla-II and its targeting platform, dynamically reconfigurable resource array (DRRA) \cite{drra_thesis, drra_interconnect} and Distributed Memory Architecture (DiMArch) \cite{dimarch_1, dimarch_2}, are one such combination that is specifically designed for the dense linear algebra application domain. To efficiently handle vector operations, DRRA has been designed as a coarse grain reconfigurable architecture (CGRA) fabric. Unique features such as the distributed two-level control (D2LC) scheme, complex address generation ability, flexible communication and clustering, etc. are included. The unique architecture of DRRA rejects any off-the-shelf synthesis tool other than Vesyla-II since Vesyla-II is dedicated to such a CGRA platform with the D2LC scheme. Fig.~\ref{fig:als} shows the relationship between Vesyla-II, DRRA and application-level synthesis (ALS). It visualizes how Vesyla-II enables the ALS for streaming applications.

The purpose of this paper is to demonstrate how Vesyla-II support such a CGRA platform with the D2LC scheme by focusing on its unique intermediate representations (IRs). The rest part of the paper is organized as follows: Section II introduces background knowledge about the DRRA platform and serves as a state-of-art review by comparing Vesyla-II with conventional HLS tools. Section III explains in detail how Vesyla-II works by focusing on its internal data structure. And finally, section IV concludes the paper and points out the direction for future works.

\section{Background}
\label{sec:background}

\subsection{Synchoros VLSI Design and Its Synthesis Flow}
In this sub-section, we provide the background information on the VLSI design style that Vesyla targets which differentiate it from the standard cell based VLSI design style targeted by conventional high-level synthesis (HLS) tools.

\begin{figure}[h]
    \centering
    \includegraphics[width=.9\textwidth]{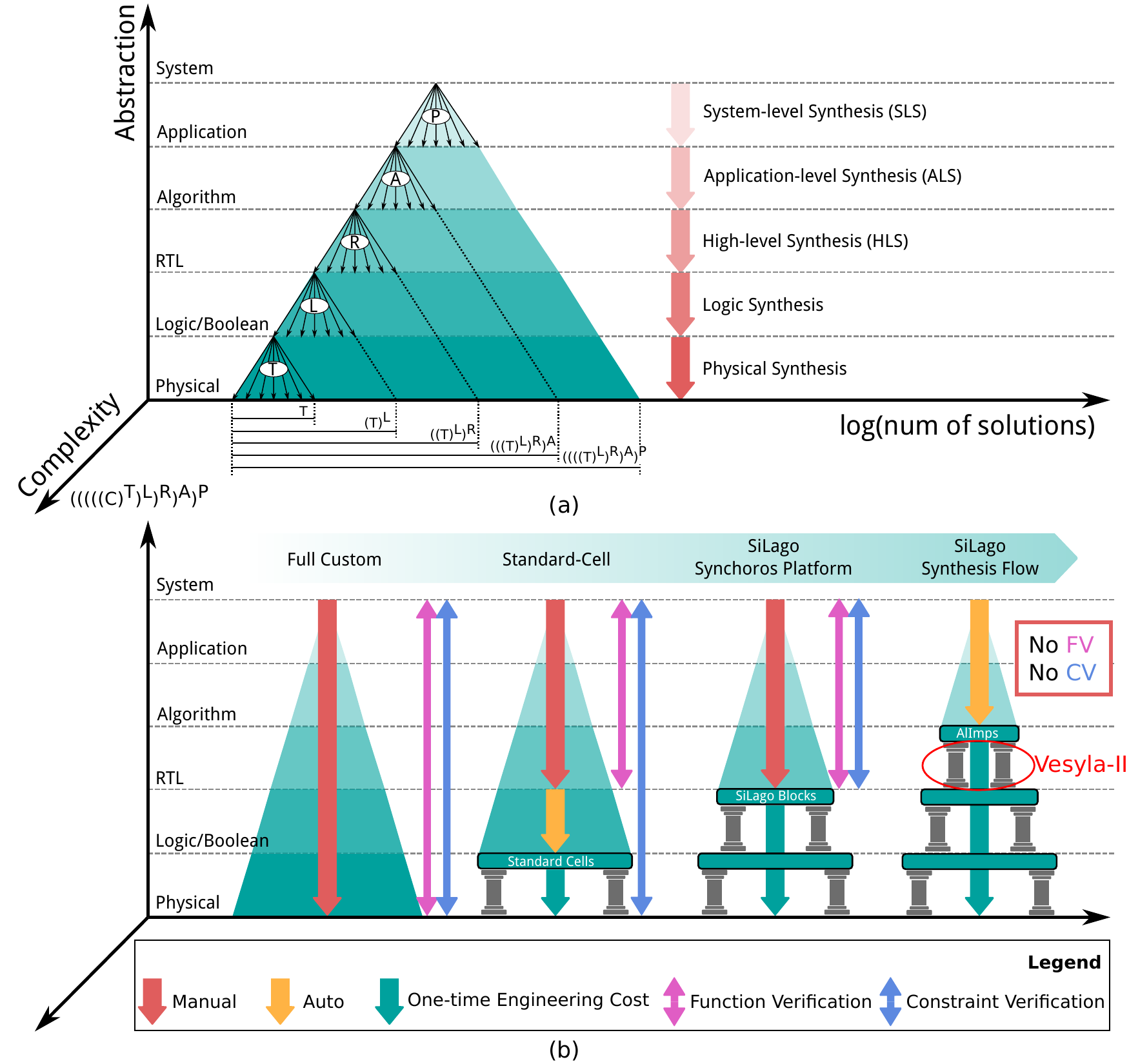}
    \caption{Synchoros VLSI Design Style. Based on \cite{Stathis2022SynchorosStyle, Hemani2017SynchoricityAffordable}}
    \label{fig:vlsi}
\end{figure}
 
VLSI design space is explored and refined successively from the highest, system-level abstraction to the lowest, physical design level abstraction. There are two challenges with this design space and its top-down refinement process also known as synthesis. The first is that the design space increases exponentially with the abstraction from which the refinement process starts as shown in Fig.~\ref{fig:vlsi}. The second is that the cost metrics of the design, the area, latency, and energy, are not known until the physical design has happened. The synthesis decisions at a higher level are made on crude estimates of what the final cost metrics will be when the physical design has happened. The accuracy of these estimates also degrades with how further the abstraction is from the physical level. This is a problem that exists even at the logic-synthesis level, which is just one abstraction high from physical design. For high-level synthesis, the estimation is exponentially worse and the design space exponentially larger. This explains, why even after three decades of intense research, HLS is still not mainstream. These challenges, as said, would become exponentially worse as we move to even higher abstractions like application and system levels.

As shown in Fig.~\ref{fig:vlsi}, the full custom design style refines the system specification and explores the design space manually. Little or no automation is involved. Therefore, it's impossible to design large chips using the full custom design style because the design space is simply too large to be explored manually. The standard cell based VLSI design style raises the abstraction from the physical level to the logic level by fixing the standard cells. The design of standard cells is a one-time engineering cost (OTEC). It can enable efficient automatic synthesis from RTL to logic level but fails to automate the complete synthesis task from system level to logic level. The synchoros VLSI design style has been proposed to get around these challenges. The word synchoros is derived from the Greek word \emph{choros} (space). In synchronous systems, time is discretised to enable temporal composition by abutment. Similarly in synchoros systems, space is discretised to enable spatial, electrical and technological composition by abutment. Domain specific coarse grain reconfigurable cells are hardened and made abutment ready. They are called SiLago (Silicon Lego) blocks. SiLago blocks occupy an integer number of grid cells and absorb all inter-block wires, including the clock, power grid, reset etc. As a result, when SiLago blocks are placed on the grid, they compose by abutment to create larger valid VLSI designs without any need for further logic or physical synthesis. See \cite{Hemani2017SynchoricityAffordable}, for more details.

The similarity between conventional HLS tools and Vesyla is that they both refine an algorithm into a set of RTL operations for datapath and control. The difference is that conventional HLS tools must estimate the cost metrics of RTL operations and follow up the HLS by logic and physical synthesis. Vesyla, in contrast, knows, not only the cost metrics of the RTL operations exported by the SiLago blocks but also the cost of the synthesised implementation of the algorithm composed in terms of the SiLago blocks. Effectively, SiLago blocks in synchoros VLSI design space raise the physical design level to RTL for both logic and wires. This has three beneficial effects. The first is that it exponentially reduces the design space that HLS tool like Vesyla needs to explore and refine, as shown in Fig. \ref{fig:vlsi}. The second is that the reduced space is composable in temporal, spatial, electrical, and technological terms. The third is that the reduced space is predictable with agile and post-layout accurate knowledge, not estimates, of the cost metrics.

Vesyla carries the spirit of synchoros design one step forward. It further reduces the design space by raising the physical design level to algorithm-level. It does this by creating a library of algorithm implementations (AlImps) in varying dimensions, degrees of parallelism and architecture styles. Cost metrics of each of these implementations are known with post layout accuracy. This significantly makes the task of application-level synthesis (ALS) simpler. Consider that an application has A algorithms, and each algorithm has R variants created by Vesyla. This creates $R^N$ possible solutions and the ALS in synchoros VLSI design space would know the cost metrics of each of these solutions with agility and post-layout accuracy. In contrast, an ALS working in the standard cell design space would need to explore an exponentially larger design space, $((T^L)^R)^A$, with crude estimates of cost metrics.

\subsection{SiLago Platform for Streaming Application}

SiLago platform is a heterogeneous platform. Different types of SiLago blocks are created to match different application domain requirements. Specifically, the two types of SiLago blocks that Vesyla targets are designed for streaming applications. One is called Dynamically Reconfigurable Resource Array (DRRA) \cite{drra_thesis, drra_interconnect} for vector variable computation, The other is called Distributed Memory Architecture (DiMArch)\cite{dimarch_1, dimarch_2} for scratchpad SRAM storage . The two fabrics are tightly coupled and can be considered composite fabrics. 

\begin{figure}[h]
    \centering
    \includegraphics[width=.9\textwidth, trim=.01cm .01cm .01cm .01cm, clip]{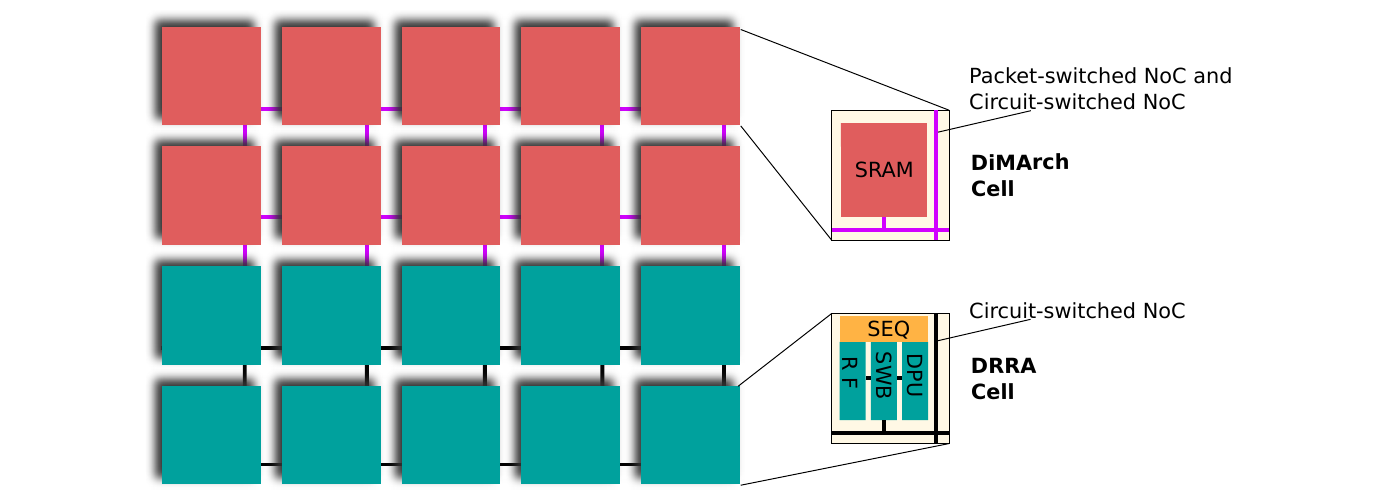}
    \caption{SiLago Platform for Streaming Application. It consists of a DiMArch fabric (red) and a DRRA fabric (green). Each cells are connected with its neighbors via different NoC systems.}
    \label{fig:silago}
\end{figure}

The architecture of DRRA and DiMArch fabrics is demonstrated in Fig.~\ref{fig:silago}. Both DRRA and DiMArch fabrics are CGRAs. All these cells are interconnected via circuit-switch or packet-switch NoC. In each DRRA cell, there is a Register File (RF) for local storage, a Datapath Unit (DPU) for arithmetic computation, a Switchbox (SWB) for interconnection configuration, and a Sequencer (SEQ) working as a controller. In each DiMArch cell, there is an SRAM block that is used as a scratchpad memory. Some highlights of DRRA/DiMArch architecture include: 1) The Address Generation Units (AGUs) in RF and DiMArch cells are highly configurable and support up to two-level affine functions. 2) The circuit switch NoC connecting DRRA cells makes it possible for a cell to access its neighbours' resources without timing penalty. The overlapped resource sharing region can span up to 5 columns. 3) The RFs and DPUs can be chained to create arbitrary complex datapath via SWB configuration. 4) The NoC in DRRA and DiMArch guarantees the easy tweak of fabric size to match the application. 5) The DRRA cell separates address computation and address constraint computation. Address computation is done by the AGUs attached to each RF port, but address constraints are computed by the Run-time Address Constraint Computation Unit (RACCU) situated inside SEQ. See section \ref{sec:ir1}.

\begin{figure}[h]
    \centering
    \includegraphics[width=.9\textwidth, trim=.01cm .01cm .01cm .01cm, clip]{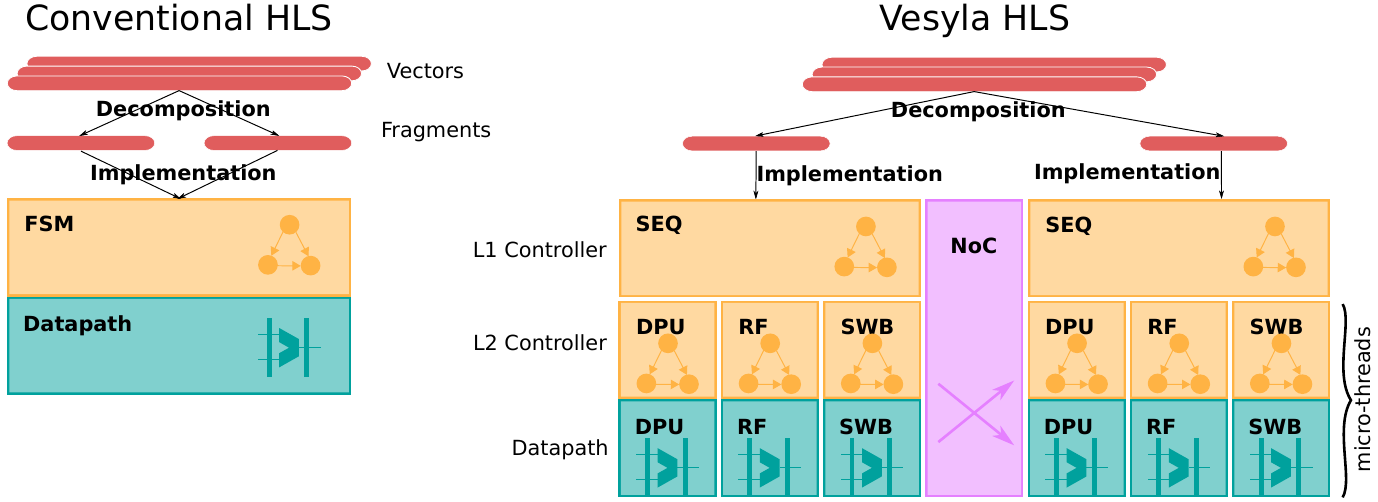}
    \caption{Comparison of conventional HLS that targets single FSM control system and Vesyla HLS that targets D2LC system. Only three types of micro-threads are shown in each DRRA cell, but in reality there are more.}
    \label{fig:d2lc_drra}
\end{figure}

Conceptually, the DRRA fabric is a vector computation machine. It can handle multiple vectors in parallel to implement arbitrary dense linear algebraic algorithms. It can further increase parallelism by breaking down each vector into fragments and processing them in parallel. The process of each vector fragment is further partitioned into multiple micro-threads. Such control scheme is supported by a Distributed Two-Level Control (D2LC) system \cite{Yang2021SchedulingInstructions, Yang2022ReducingSystem} as demonstrated in Fig. \ref{fig:d2lc_drra}. Each DRRA cell has a level-1 controller (SEQ) that sends instructions to its local level-2 controller to initiate micro-threads. Once initiated, these micro-threads operate independently by the local level-2 controllers. Typically, micro-threads are used to implement a single aspect of a vector computation, such as constructing interconnection channels, computing address or address constraint, or performing arithmetic operations on vector variables. In Fig.~\ref{fig:d2lc_drra}, only three types of micro-threads are shown, but DRRA has more types of micro-threads in reality. The D2LC system clearly sets DRRA apart from other state-of-the-art CGRA accelerators.
 
The intermediate representations and the synthesis algorithms of Vesyla-II differ from conventional HLS tools because of the different computation models, micro-architecture and the instruction-set of DRRA. In section \ref{sec:vesyla}, we highlight how Vesyla-II supports the DRRA architecture by explaining the intermediate representation of each compilation stage.

\subsection{State-of-the-Art}

Conventional HLS tools have been researched for decades and some of them have already been commercialized. Stratus \cite{CadenceSynthesis}, Synphony C \cite{SynopsysSynthesis}, and Vivado HLS \cite{Feist2012VivadoSuite} are all successful commercial HLS products. Academic HLS tool like GAUT \cite{Coussy2008GAUTApplications} also contributes to the HLS community. Despite the initial success of HLS, all the tools generate unstructured RTL code with a single control unit. Iterative logic synthesis and placement \& routing steps are still needed to reach GDSII physical level. These commercial or academic tools cannot be used directly to synthesize programs for DRRA because the DRRA has a unique D2LC system which is not supported by any of these tools.

Even though we position Vesyla-II to be an HLS tool in the SiLago synthesis flow, it can also be viewed as a CGRA compiler. Many compilers have already been created for various CGRA platforms, such as DRESC for ADRES \cite{Mei2008ADRESProcessors}, the RaPiD-C compiler \cite{CronquistSpecifyingRaPiD}, the compiler for Montium processor \cite{HeystersMappingArchitecture}, the compiler for PipeRench \cite{Budiu1999FastFabrics}, and the XPP-VC compiler for the XPP architecture \cite{Cardoso2002XPP-VCArchitecture, Cardoso2003DesignExhibition} etc. Some of the CGRA targets accelerate the innermost loop and use DFG to represent the algorithm. Some target more complex algorithms and allow loop and branch. But none of them can support the distributed 2-level control (D2LC) system like the DRRA. Scheduling and synchronization become more complex when dealing with the D2LC system.

The centrepiece of compiler design is the design of intermediate representation (IR). A good IR can expose the right amount of information that can easily bridge the input programming language and targeted machine. A popular list-based IR is the LLVM IR. The LLVM \cite{LattnerLLVM:Transformation} is an actively developed compiler framework with a large community. But Vesyla-II can't reuse its code because 1) the IR of LLVM is mainly designed for scalar operations, even with the extended vector support such as \cite{Stephens2017TheExtension}, it's still not natural to express the operations on vector machines, e.g. the LLVM IR lack of a flexible way to express the rich range addressing mode. 2) The back-end of LLVM is platform-dependent, since Vesyla-II targets on SiLago platform, we still have to develop a new back-end. Hence, LLVM is not very useful for the development of Vesyla-II. Tree-based data structure like the Abstract Syntax Tree (AST) is good to represent the input language but it's not very good to express the concurrent execution. Therefore, it's not very common to have AST as the backbone IR of complex hardware synthesis tools. The Dataflow graph (DFG) and Control/Dataflow Graph (CDFG) \cite{Gajski1992DesignTransformations, Cooper2012EngineeringCompiler} are the most common IR used both in HLS tools and CGRA compilers because such graph-based IR can naturally capture the massively parallel execution on the hardware platform. The dependency graph \cite{Cooper2012EngineeringCompiler} is also used in some compilers and synthesis tools to describe the relative order of each operation in the graph. In this paper, we will focus on demonstrating the IRs used in Vesyla-II. Different from the conventional compiler design, Vesyla-II uses multiple IRs. Each of which serves the data structure foundation for various optimization processes. The IRs in Vesyla-II are also graph-based. Most of them are an extension of the existed data structure like the CDFG.

This paper describes the second version of Vesyla. The Vesyla-I (see \cite{vesyla1, vesyla2, algosil}) didn't use well designed IR. Both the front-end and back-end were based on the AST. It was extremely difficult to exploit the instruction level parallelism. There were not many optimization processes used in Vesyla-I except for some well studied general optimization techniques, such as constant folding, and some ad-hoc platform-specific optimizations. The efficiency of the instruction scheduling process was affected most by the AST-based IR because the atomic operation modelled by such IR was a complete vector operation which was simply too coarse grain to be exploited for micro-thread level concurrency. In short, the old version of Vesyla focused more on converting the algorithm from high-level language to the DRRA and DiMArch mapping other than on trying to optimize the mapped result.
\section{Vesyla-II}
\label{sec:vesyla}

In this section, we elaborate on the construction of Vesyla-II, focusing on its processing steps and data structures. We justify our design policies and highlight how Vesyla-II support the D2LC scheme and other important features of the DRRA vector machine such as its address generation mechanism.

\subsection{Overview of Synthesis Steps}

\begin{figure}[h]
    \centering
    \includegraphics[width=.9\textwidth, trim=.01cm .01cm .01cm .01cm, clip]{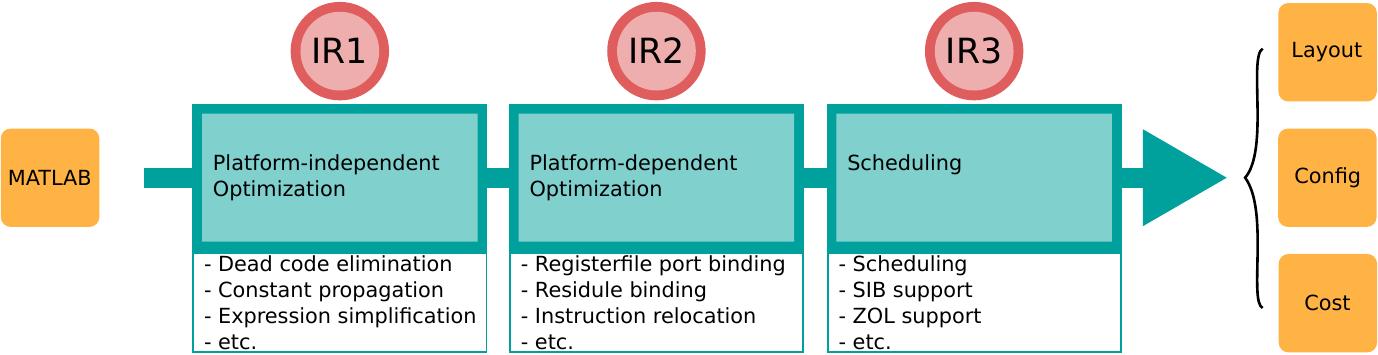}
    \caption{The Overview of Vesyla-II Synthesis Process and Data Structure.}
    \label{fig:vesyla}
\end{figure}	 

The high-level organization of Vesyla-II is illustrated in Fig.~\ref{fig:vesyla}. Vesyla-II accepts a subset of MATLAB code with pragmas and emits the layout of DRRA and DiMArch fabric, the configuration for each distributed controller and the estimated cost of the synthesized algorithm. In the figure, three important data structures have been marked out by red circles. They are the intermediate representations (IRs) of Vesyla-II in its different synthesis stages. All of them are graph-based data structures:

\begin{itemize}
    \item IR1: Control address dataflow graph (CADFG)
    \item IR2: Instruction dependency graph (IDG)
    \item IR3: Hierarchical multi-thread dependency graph (HMTDG)
\end{itemize}

The IR1 is the data structure that captures the meaning of the input MATLAB program. It models the data and control flow by emphasizing the stream-like vector data movement and its corresponding address generation. The IR1 serves as the data structure that is used for operation level platform independent transformation and optimization. The IR2 is the data structure that captures the mapping on the SiLago platform. It expresses the relationship of each instruction that will execute on the D2LC controller. The IR2 serves as the foundation of instruction level platform-dependent transformation and optimization. The IR3 is the data structure that illustrates the detailed timing relation of each instruction during its lifetime. Instructions of the DRRA platform tend to have a long lifetime and are highly cooperative (see \cite{Yang2021SchedulingInstructions}). Each instruction needs to be decomposed into even smaller units called instruction phases in IR3. The IR3 serves as the data structure that enables the detailed scheduling of these instructions.

\subsection{Input Format and Pragma Guided Binding}

Ideally, HLS tools are expected to find near optimal solutions from a pure untimed functional model of the algorithm. However, the design space of the HLS is very large, and most HLS tools allow end-users to use pragmas to guide themselves to reach the desired architecture. In the same spirit, Vesyla-II also supports a set of \emph{pragmas} to guide it towards desired architecture. As stated, the role of Vesyla-II is not just a pure HLS tool. It is a library development tool that enables the application-level synthesis (ALS). The pragma used in Vesyla-II can also be symbolic and written in the Jinja language format \cite{jinja}. Symbolic parameters are ideal for sweeping for the desired range by the library developer or ALS tools for each algorithm to generate a set of Pareto front points that vary in terms of their cost metrics: area, latency and average energy.

\begin{figure}[h]
    \centering
    \includegraphics[width=.9\textwidth, trim=.01cm .01cm .01cm .01cm, clip]{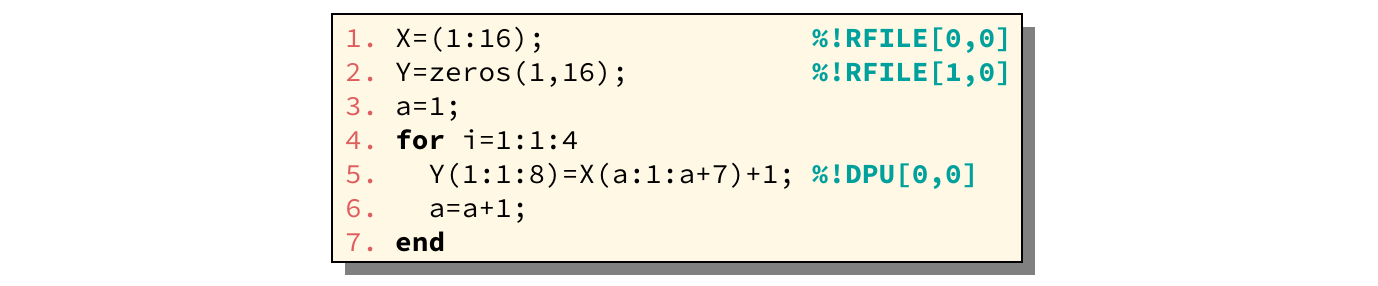}
    \caption{Example Matlab Program Accepted by Vesyla-II.}
    \label{fig:input}
\end{figure}

The pragmas supported by Vesyla-II are mainly used for guided binding of memory storage and major arithmetic computation units. An example is shown in Fig.~\ref{fig:input}. In this simple example, we assign variables to either register files in a different DRRA cell. It is also possible for multiple variables to be assigned to the same SRAM bank or Register files. In such a situation, Vesyla-II takes care of address transformation, time multiplexing and scheduling of ports and paths to SRAM and register files.

Even though deciding on pragmas are manual effort, they do not compromise the spirit of automation. The pragmas only guide Vesyla-II with some critical aspects of and binding that acts more like constraint specifications. The more cumbersome and error prone aspects of synthesis, like detailed scheduling, synchronization and generation of configuration instructions embodying the multiple concurrent FSMs, are fully automated. It is not an extremely complex task to automate the major resource binding. A heuristic algorithm such as resource constrained coarse grained LIST scheduling that optimizes the latency can accomplish such a task. However, the binding problem for each algorithmic function is small and straightforward in the eyes of an experienced library developer. Maintaining such a pragmatic interface would be very convenient for them to interfere with the synthesis process of Vesyla-II.

Another important aspect of the synthesis is resource allocation. In Vesyla-II, it is completely manual. Since Vesyla-II is a library development tool, it needs to generate multiple solutions with different cost metrics from the same input function. Their variation directly comes from their resource budgets. Therefore, Vesyla-II should support such an interface to allow library developers or ALS tools to specify the allocation strategy instead of coming up with one by itself.

\subsection{IR1: Control Address Dataflow Graph}
\label{sec:ir1}

Conventional ALS tools typically model high-level programs by dataflow graphs (DFG) or control dataflow graphs (CDFG). Vesyla-II extended the control dataflow graph to better deal with streaming vector operations. We call it the control address dataflow graph (CADFG) and it is the first important IR in Vesyla-II synthesis steps.

\begin{figure}[h]
    \centering
    \includegraphics[width=.9\textwidth, trim=.01cm .01cm .01cm .01cm, clip]{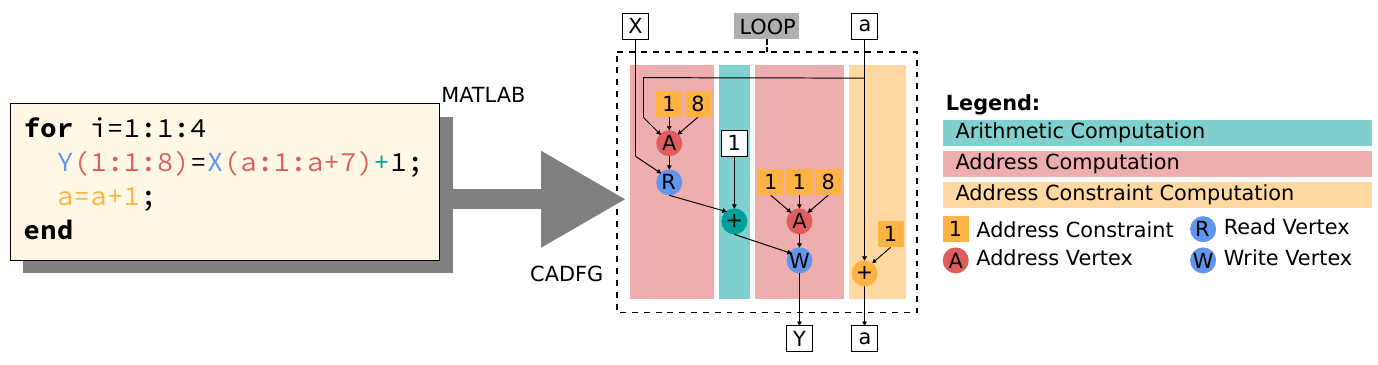}
    \caption{Address Generation in IR1 and the overlapping of micro-threads involving arithmetic computation, address computation and address constraint computation.}
    \label{fig:cadfg}
\end{figure}

The CADFG extends CDFG by adding the address generation information for vector operations. A vector operation can be decomposed into a temporal stream of scalar operations by supplying a proper accessing address at each temporal iteration. Similarly, an element-wise scalar operation wrapped by loops can also be replaced by a vector operation. To guarantee the functionalities are identical, address generation needs to be properly synthesized. Vesyla-II supports synthesizing the maximum two-level affine function as the address generation function for each vector element access. We choose the affine function as the supported format because, in dense linear algebra, most indexing functions are affine functions. Figure~\ref{fig:cadfg} illustrates how an address generation for vector operation is represented in CADFG. Here, all the address generation functions are simple one-level affine functions.

In the example shown in Fig.~\ref{fig:cadfg}, the slicing operation in the MATLAB code is translated to the \verb|A| vertex in the CADFG. The \verb|A| node is called the \emph{address vertex}. In Fig.~\ref{fig:cadfg}, we only use one-level affine function as the address generation function. Each address vertex receives \emph{address constraints} and generate a stream of address that will be used by the \emph{read} or \emph{write} vertex. The address constraints are not always constants. It might be necessary to compute them at run-time. These address constraint computations are usually scalar operations and are computed at each iteration of loops. The DRRA naturally supports address generation and address constraint generation. Separating those two from arithmetic computation prevents the potential interruption of the arithmetic computation. They are designed to work in separate micro-threads implemented by the level-2 controller. This is also the reason why DRRA architecture has a two-level control scheme. Each micro-thread of the example code is highlighted in Fig.~\ref{fig:cadfg}. In each iteration of a loop, address constraints are computed for the next loop iteration so that it does not delay the program execution.

In terms of control structure, Vesyla-II supports the for-loops and branches. The for-loop supported by Vesyla-II can be parametrically static. As long as the loop bound does not rely on the input data, Vesyla-II can handle it regardless of whether it needs run-time computation or not. The while-loop is not supported because its high degree of dynamism is not directly supported by the DRRA platform and it’s impossible to schedule the execution of instructions inside such a dynamic region at compile-time. Both the for-loops and branches of MATLAB are directly represented by a \verb|LOOP| vertex and a \verb|BRANCH| vertex in IR1. To use only a single vertex to represent these control concepts makes the IR1 simple and easy to understand. It also makes the conversion from IR1 to IR2 easier because the targeted platform of Vesyla-II has dedicated infrastructures to support loops and branches.

When building IR1, Vesyla-II also records data hazards that will be gradually refined. Three kinds of data hazards are recorded in IR1: Read-After-Write (RAW), Write-After-Read (WAR) and Write-After-Write (WAW). The simple RAW hazard will be directly absorbed by the data dependency edge that is anyway created. The WAR and WAW require the creation of additional special dependency edges to indicate that there are data dependencies. Data hazards for vector variables are more difficult to handle compared to scalar variables because we have to analyze not only the variable name but also the access pattern.

Data dependency edges indicating data hazards are created by Vesyla-II while building IR1 without analyzing them. It faithfully records all data dependencies without discrimination. During the dependency analysis phase, Vesyla-II will then analyze the data access pattern of each dependency edge and categories them into proper types.

Data dependencies between vector variables are more complex than scalar variables because it not only involves the variables but also their read and write access patterns. There are many sophisticated techniques to analyze dependencies among vector operations. Many great works have been done in this research field \cite{Allen1987AutomaticForm}. However, Vesyla-II, as a proof-of-concept tool, does not explore all those advanced methods for optimization purposes. It broadly categorizes data dependency among vector variables as 1) strong dependency 2) weak dependency and 3) fake dependency. Strong dependency requires the successor to start no earlier than the end of its predecessor. Weak dependency only requires the successor to start no earlier than the start of its predecessor. Fake dependency indicates there is no actual data hazard between predecessor and successor, the dependency edge can be removed.

\begin{figure}[h]
    \centering
    \includegraphics[width=.9\textwidth, trim=.01cm .01cm .01cm .01cm, clip]{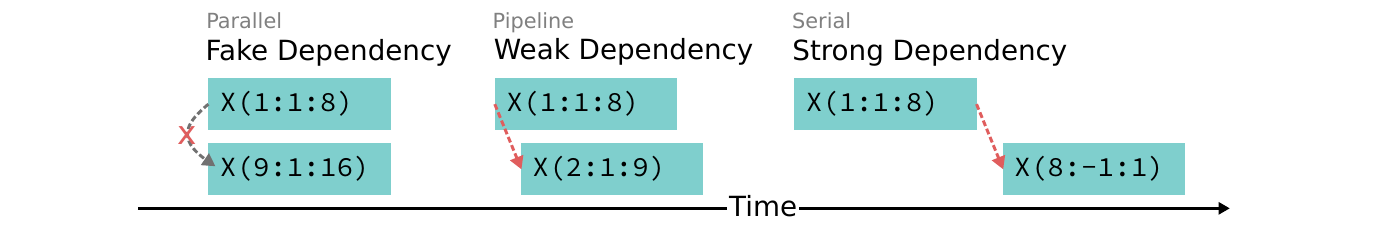}
    \caption{Different Categories of Data Dependency.}
    \label{fig:hazard}
\end{figure}

As shown in Fig.~\ref{fig:hazard}, If the predecessor and successor of a dependency edge have non-overlapping access patterns the dependency edge is fake because the two nodes will never access the same location. If the predecessor and successor have the same pattern or the successor has a pattern that is a simple shift of the predecessor pattern, the dependency is weak because the predecessor will always access the location before the successor, thus the successor can start one cycle later than the starting of the predecessor. For all other cases, Vesyla-II considers the dependency edge strong since this is the most conservative assumption.

Besides data access pattern analysis, Vesyla-II performs a series of other platform-independent transformations to optimize the input MATLAB code based on IR1. The constant folding, constant propagation and dead code elimination are implemented to simplify the arithmetic operations. Currently, Vesyla-II does not perform more advanced transformations such as loop strip-mining and loop unrolling \cite{Kennedy2001OptimizingApproach}, but we have plans to add them in the future.

\subsection{IR2: Instruction Dependency Graph}

As the basis of platform-dependent transformation and optimization, the Instruction Dependency Graph, which is the second intermediate representation in Vesyla-II (IR2), shows the coarse-grain relationship of each instruction that will be executed on the targeted platform. In this section, we first define the instruction set and explain how the instructions execute on the DRRA platform. Then we demonstrate an example of IR2 and explain the conversion from IR1 to IR2 for the example.

We first introduce the instruction set that is supported by DRRA CGRA fabric. It consists of 9 instructions divided into five categories:

\begin{itemize}
    \item Computation instructions (C)
    \item Address generation instructions (A)
    \item Address constraint generation instructions (AC)
    \item Interconnection instructions (I)
    \item Control instructions (CTR)
\end{itemize}

Table \ref{tab:isa} lists all the instructions that are supported by the DRRA platform. Other similar platforms that have distributed two-level control systems could use the same instruction set or its extended version. The instruction set does not tide to a specific hardware implementation.

\begin{table}[H]
\caption{Instruction Set of DRRA}
\label{tab:isa}
\begin{tabular}{ |l|l|l| }
 \hline
 Instruction & Category & Description \\ 
 \hline
DPU & C & Configure computation unit\\
REFI & A & Configure the register file port\\
SRAM & A & Configure the scratchpad memory port\\
RACCU & AC & Compute an address constraint\\
SWB & I & Configure a data transfer path on DRRA\\
ROUTE & I & Configure a data transfer path on DiMArch\\
LOOP & CTR & Configure a loop controller\\
BRANCH & CTR & Branch according to computation flag\\
JUMP & CTR & Unconditional jump\\
WAIT & CTR & Do nothing, used for synchronization\\
 \hline
\end{tabular}
\end{table}

The DRRA instructions are resource-centric instructions which are very similar to instructions in the No Instruction-Set Computing (NISC) paradigm \cite{Reshadi2005ADatapaths}. The DRRA uses a distributed two-level control (D2LC) system. Instructions of such a control system are persistent and fully cooperative \cite{Yang2021SchedulingInstructions, Yang2022ReducingSystem}. Every instruction listed in Table \ref{tab:isa} is fetched and issued by the level-1 controller and is used to configure a level-2 controller. Even though some of the level-2 controllers are very simple, single-state FSMs, and in practice are merged into the level-1 controller, conceptually we still consider them as fully functional level-2 FSMs. Fig. \ref{fig:d2lc_drra} shows a simplified DRRA fabric which highlights the relationship among the level-1 controller, the level-2 controllers and the datapaths. Only three datapaths are shown in each DRRA cell.


\begin{figure}[h]
    \centering
    \includegraphics[width=.9\textwidth, trim=.01cm .01cm .01cm .01cm, clip]{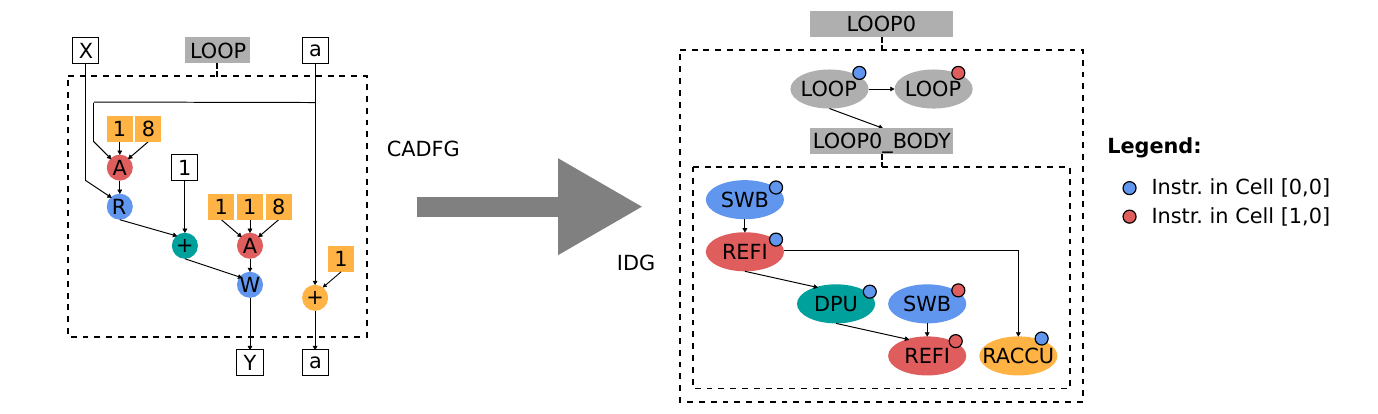}
    \caption{The IDG converted from the CADFG.}
    \label{fig:idg}
\end{figure}

Fig.~\ref{fig:idg}. demonstrates how Vesyla-II synthesizes instruction vertices in IR2. For example, the \verb|LOOP| vertex in IR1 is replaced by a two-level hierarchical loop structure. the \verb|LOOP0| represent the complete loop. Two \verb|LOOP| instructions are synthesized for each DRRA cell. A \verb|LOOP0_BODY| vertex is synthesized to host all internal instructions of the loop body. Inside the loop body, the computation vertex in IR1 has also been replaced by a \verb|DPU| instruction vertex with proper configuration mode to guarantee that the computation unit is configured to perform the same arithmetic function. Address generation is similarly replaced by \verb|REFI| instructions and address constraint generation is replaced by \verb|RACCU| instruction vertices. Notably, the edges between address generation and computation also result in a \verb|SWB| instruction vertex in IR2, that is because in IR1 we don’t consider the construction of data transfer passage, while in IR2, according to the specification of the platform, we have to actively build the data transfer channel by interconnection instructions such as \verb|SWB| instruction. In Fig.~\ref{fig:idg}, we have removed the detailed weight information associated with each edge to avoid cluttering the graph. In reality, each edge specifies how exactly its source and destination vertices are temporally related. This information will be utilized when transforming IR2 to IR3 in the next section.

Based on IR2, Vesyla-II gradually expand each statement in CADFG to pre-defined instruction macros and the properties of each instruction vertex are properly configured based on the information in the IR1. Several instruction-level transformations and optimization are also been implemented in this stage. For example, a certain degree of code motion is automatically exploited by the natural parallel graph structure. The operator strength reduction and tree height reduction are implemented to map the address constraint computation on very restricted hardware resources. Automatic scalar register allocation and binding for address constraint computation are also performed by Vesyla-II to minimize the address constraint computation latency. More information on these standard compiler optimization processes can be found in \cite{Muchnick1998AdvancedImplementation}. Besides these well-studied optimization techniques, Vesyla-II also performs an optimization process that targets specific features on DRRA. For example, Vesyla-II will try to move the \verb|SWB| and \verb|ROUTE| instructions outside the loop to avoid unnecessary path reconfiguration. Similarly, \verb|REFI| and \verb|SRAM| instructions are also moved outside the loop if possible to in-cooperate repetitive address access pattern to a single \verb|REFI| or \verb|SRAM| instruction.

\subsection{IR3: Hierarchical Multi-Thread Dependency Graph}

The IR2 models the original algorithm in terms of the instructions supported by DRRA. However, the spatio-temporal details of those instructions are not expressed by IR2. Further refinement on IR2 in terms of scheduling and synchronization is not possible. For this reason, we further refine instructions in terms of their phases of execution. The refined data structure then becomes the third intermediate representation of Vesyla-II (IR3) and it is called a Hierarchical Multi-Thread Dependency Graph (HMTDG).

The instruction set used by DRRA is very different from the traditional instruction set of microprocessors. These instructions on DRRA are implemented by individual level-2 FSMs that can reside for a very long time once configured. Each instruction by itself cannot accomplish a complete computation task, they have to cooperate with other instructions. We call such instructions \emph{persistent and fully cooperative instructions} \cite{Yang2021SchedulingInstructions}. These instructions demand a new data structure that can capture the fine-grained timing and resource hazards. The HMTDG can handle such a task because it models more spatio-temporal details for the components and phases of the instructions that must be refined to be able to do scheduling and it serves to abstract away the implementation details of the DRRA instructions.

A typical streaming application program usually has a hierarchical structure. The top-level container would be the whole program, in which inner loop structures and/or branch structures can reside. HMTDG is a hierarchical graph structure in which some vertices could embed child graphs. These hierarchical vertices implement control structures like loops and branches.

\begin{figure}[h]
    \centering
    \includegraphics[width=\textwidth]{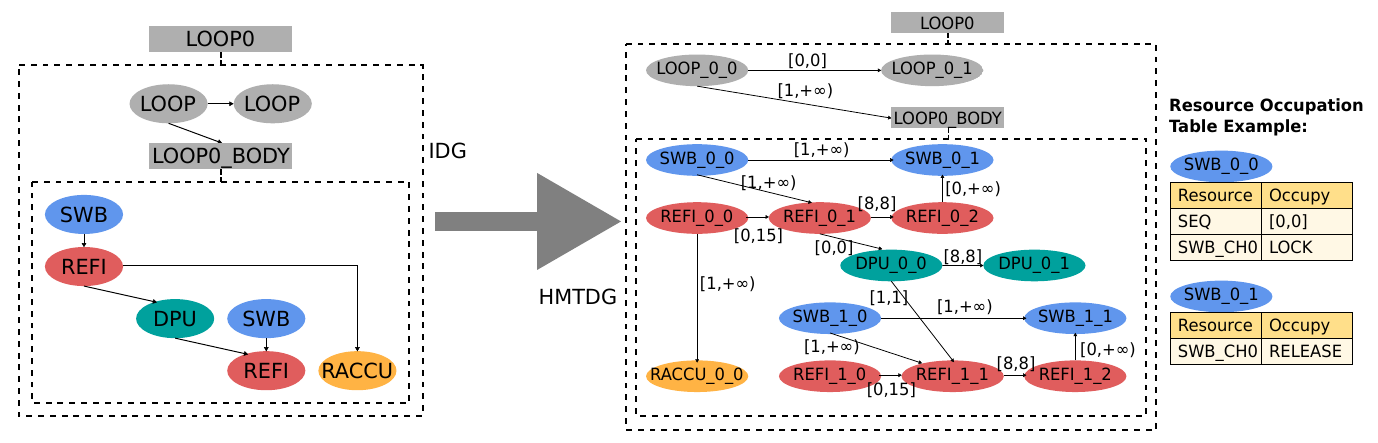}
    \caption{The HMTDG converted from the IDG}
    \label{fig:hmtdg}
\end{figure}

Fig.~\ref{fig:hmtdg} demonstrates how an IDG is converted to a HMTDG. The vertices in both IDG and HMTDG are colour-coded. We can see that each instruction in IDG is decomposed to one or more nodes in HMTDG. We call the vertex in HMTDG the \emph{instruction phases}. We know that the instructions running on DRRA are persistent. Each instruction is used to configure a level-2 FSM. Therefore, each instruction phase corresponds to each critical state transition of the level-2 FSM. For example, the \verb|REFI| instruction is translated to three phases: \verb|Fetch/Issue|, \verb|activation|, and \verb|end|. The dependency arrow between vertices in IDG is refined to arrows in HMTDG. The weight on each arrow indicates the time constraint of the two vertices. Any arrow $A \xrightarrow{[m, n]} B$ indicates that the node $B$ should be scheduled after node $A$ at least $m$ cycles and at most $n$ cycles. The vertices in the HMTDG also have a resource occupation table. In the figure, there are two examples of such a table. For example, the \verb|SWB_0_0| will occupy the \verb|SEQ| resource for 1 cycle and it will \verb|LOCK| the \verb|SWB_CH0| resource until some other vertex release the resource in the future. The occupation time in the resource occupation table is always relative and uses the scheduled time of the vertex as the reference point.

On the leaf level, HMTDG will be a single plain graph which can be converted to a data structure called \emph{Persistent and Fully Cooperative Instruction Model (PCIM)}. Such data structure can be scheduled by the algorithms presented in \cite{Yang2021SchedulingInstructions}. A scheduling task for the whole HMTDG can be decomposed into many smaller scheduling tasks that work on a single PCIM. Once the leaf PCIMs are scheduled, we can reduce one hierarchical level of the original HMTDG. Thus, step by step, the whole HMTDG can be scheduled.

After the scheduling phase, instruction synchronization is performed to insert \verb|WAIT| instructions and finally determine the PC value of each branching point. After instruction synchronization, Vesyla-II generates a collection of files, including an instruction list (binary and verbose), DRRA and DiMArch layout, RTL testbench and profiler, and several scripts that accelerate the functional test process.
\section{Conclusion and Future Wroks}
\label{sec:fw}

In this paper, we explained the necessity of adopting the synchoros VLSI design style for developing an efficient complex hardware system. As part of the synchoros VLSI design flow, the HLS tool Vesyla-II has been designed for synthesizing algorithms based on synchoros SiLago blocks. Compared to the Vesyla-I, we have redesigned the previous ad-hoc intermediate representations with data structures that can address the different aspects of both vector operations and the distributed two-level control (D2LC) system. 

The next step is to further enhance the Vesyla-II to improve the quality of the generated designs. It includes a better address pattern analysis mechanism, more powerful instruction rearrangement algorithms, and a more efficient instruction scheduling algorithm. We also need to improve the Vesyla-II by adding an automatic binding process to make it more user friendly.

As a library development tool, we need to use Vesyla-II to populate the common libraries for streaming applications. Such libraries could include linear algebra libraries, signal processing libraries, artificial neural networks, etc.

To complete the whole synthesis flow, application-level synthesis (ALS) needs more attention. The ALS tool should be able to synthesize applications based on the algorithm library created by Vesyla-II. It needs to explore the design space and synthesize each algorithm instance, the interconnection, the memory hierarchy, and the global control system. The floor plan of the whole application should also be taken care of by the ALS tool.

\bibliographystyle{ACM-Reference-Format}
\bibliography{contents/reference.bib}

\end{document}